\documentclass[aps,pra,twocolumn,amsmath,amssymb,footinbib,showpacs]{revtex4}
\newcommand{\Jnature}{Nature (London)}
\newcommand{\Jnatphys}{Nature Phys.}

\newcommand{\Jprl}{Phys. Rev. Lett.}
\newcommand{\Jpr}{Phys. Rev.}
\newcommand{\Jpra}{Phys. Rev. A}
\newcommand{\Jprb}{Phys. Rev. B}

\newcommand{\Jrmp}{Rev. Mod. Phys.}

\newcommand{\Jnjp}{New J. Phys.}

\newcommand{\Jchinpl}{Chinese Phys. Lett.}

\newcommand{\Jjphyschemsol}{J. Phys. Chem. Sol.}

\newcommand{\JjphysB}{J. Phys. B: At. Mol. Opt. Phys.}

\newcommand{\JZphys}{Z. Phys.}

\usepackage[english]{babel}
\usepackage{latexsym}
\usepackage{graphics}
\usepackage{subfigure}
\usepackage{epsfig}
\usepackage{color}
\usepackage{hyperref}
\hypersetup{
colorlinks=true,
citecolor=blue,
linkcolor=red,
urlcolor=black
}

\renewcommand{\i}{\textrm{i}}

\newcommand{\be}{\begin{equation}}
\newcommand{\ba}{\begin{align}}
\newcommand{\ee}{\end{equation}}
\newcommand{\ea}{\end{align}}

\newcommand{\Ham}{\mathcal{H}}
\newcommand{\Su}{S^u}
\newcommand{\Sx}{S^x}
\newcommand{\Sy}{S^y}
\newcommand{\Sz}{S^z}
\newcommand{\Spx}{S'^x}
\newcommand{\Spy}{S'^y}
\newcommand{\Spz}{S'^z}
\renewcommand{\vec}[1]{\mathbf{ #1}}

\newcommand{\Jo}{J_\textbf{\tiny 0}}

\begin{document}

\title{Quantum magnetism of bosons with synthetic gauge fields
in one-dimensional optical lattices: a Density Matrix Renormalization Group study}

\author{Marie Piraud$^1$}
\author{Zi Cai$^1$}
\author{Ian P. McCulloch$^2$}
\author{Ulrich Schollw\"ock$^1$}
\affiliation{
$^1$~Fakult\"at f\"ur Physik,
LMU M\"unchen,
Theresienstrasse 37,
D-80333 M\"unchen, Germany\\
$^2$~Centre for Engineered Quantum Systems,
School of Physical Sciences,
The University of Queensland,
Brisbane, Queensland 4072, Australia}

\date{\today}

\begin{abstract}
In this paper, we provide a comprehensive study of the quantum
magnetism in the Mott insulating phases of the 1D Bose-Hubbard model
with abelian or non-abelian synthetic gauge fields, using the Density
Matrix Renormalization Group (DMRG) method. We focus on the
interplay between the synthetic gauge field and the asymmetry of the
interactions, which give rise to a very general effective magnetic
model: a XYZ model with various Dzyaloshinskii-Moriya (DM) interactions. The properties of the
different quantum magnetic phases and phases transitions of this
model are investigated.
\end{abstract}

%
%67.85.Fg (ou 03.75.Mn) Multicomponent condensates; spinor condensates
%05.30.Rt Quantum phase transitions
%75.10.Pq 	Spin chain models

%
%05.30.Jp Boson systems
%64.70.Tg Quantum phase transitions in specific phase transitions
%73.43.Nq Quantum phase transitions in Quantum Hall effects
%67.85.-d Ultracold gases, trapped gases
%67.85.Bc Static properties of condensates
%67.85.De Dynamic properties of condensates; excitations, and superfluid flow
%71.10.-w Theories and models of many-electron systems
%71.10.Fd Lattice fermion models (Hubbard model, etc.)
%71.70.Ej 	Spin-orbit coupling, Zeeman and Stark splitting, Jahn-Teller effect
%71.27.+a Strongly correlated electron systems; heavy fermions
%75.10.-b General theory and models of magnetic ordering%05.50.+q Lattice theory and statistics (Ising, Potts, etc.)
%75.10.Jm Quantized spin models, including quantum spin frustration

\maketitle

%%%%%%%%%%%%%%%%%%%%%%%%%%%%%%%%%%%%%%%%%%%%%%%%%%%%%%%%%%%%%%%%%%%%%%

\section{Introduction}

Recently, significant effort has been devoted to the realization  of
synthetic gauge fields for electrically neutral
atoms~\cite{jaksch2003,dalibard2011,galitski2013}. By suitably
coupling the atoms to laser fields, experimentalists have
successfully created both abelian (effective magnetic
fields~\cite{lin2009,aidelsburger2011}) and non-abelian gauge
potentials (effective spin-orbit coupling ~\cite{lin2011}) in
ultracold atomic systems, where the neutral atoms subjected to
synthetic gauge fields exhibit a variety of interesting phenomena,
including the Hofstadter fractal
spectrum~\cite{hofstadter1976,aidelsburger2013,miyake2013},
spin-orbit coupled Bose-Einstein condensates
~\cite{lin2011,wu2008,stanescu2008,wang2010,grass2011,ho2011,hu2012,zhang2012,zhou2013},
as well as spin-orbit coupled degenerate Fermi
gases~\cite{wang2012,cheuk2012,kennedy2013}. While most of these
studies focus on the weakly interacting regime,
the addition of a tunable optical lattice
%, with the advantage of high-tunability, 
enables us to investigate the strongly correlated Mott
insulating phases in the presence of gauge fields, where the
interplay between strong interactions and synthetic gauge fields
can give rise to exotic quantum magnetism that is difficult to access
in solid state
physics~\cite{pesin2010,rachel2010,cocks2012,cai2012,radic2012,cole2012,mandal2012,gong2012,orth2013,zhao2013}.

When the optical lattice is sufficiently deep to drive the system
%deep enough and the system is driven
into the Mott insulating phase,  the charge fluctuations are
suppressed and the physics can be captured by an effective magnetic
superexchange model. In the absence of a synthetic gauge field, it is
well known that the effective Hamiltonian is described by an
anisotropic Heisenberg model (XXZ model)~\cite{duan2003,kuklov2003},
where the anisotropy is determined by the asymmetry of the
interactions in spin or quasi-spin space (the ratio between the
inter-species and intra-species interaction strength). Introducing
synthetic gauge fields into the Mott insulating phases, as we will
show below, gives rise to a Dzyaloshinskii-Moriya (DM)
interaction ~\cite{dzyaloshinsky1958,moriya1960}, which is strongly
reminiscent of its counterpart in strongly correlated electronic
materials e.g. in the cuprate superconductor YBa$_2$Cu$_3$O$_6$
~\cite{coffey1991,bonesteel1993} or in low-dimensional magnetic
materials~\cite{oshikawa1997,affleck1999}. In electronic materials,
the spin-independent interaction (Coulomb interaction) 
%gives rise to the consequence that 
causes the leading magnetic superexchange model to be
an isotropic Heisenberg model. It is known that for the 1D isotropic
Heisenberg model, the additional DM interaction can be gauged away
by performing a spin rotation~\cite{shekhtman1992}. However, for
ultracold bosons with spin degrees of freedom, the situation is
different: the inter-species and intra-species scattering lengths
can be tuned within a broad range using Feshbach
resonances~\cite{inouye1998,donley2002}.
%\new{and tuning the lattice depth ??}.
This leads to an asymmetry of the interactions as well as
an anisotropy in the Heisenberg model -- where the DM interaction can
no longer be gauged away -- and plays an important role in
determining the magnetic properties of the system.

In this paper, we provide a comprehensive analysis %understanding 
of the quantum magnetism of the Mott insulating phases of the 1D
Bose-Hubbard model with both abelian and non-abelian synthetic gauge
fields using the Density Matrix Renormalization Group (DMRG)
method~\cite{white1992,schollwock2005}. We show that the
interplay between the synthetic gauge field and the asymmetry of the
interactions gives rise to a XYZ model with different DM
interactions (with DM vectors along the $x, y$, and $z$-directions),
which is the most general form for a 1D spin-$1/2$ quantum magnetic
model with two-site nearest-neighbor interactions. We explore the
phase diagram of this model, and analyze %different
the quantum phases and phase transitions %in this model.
that this model exhibits.

%-----------------------------------------%
\begin{figure*}[ht]
\begin{center}
\includegraphics[width=1.0\textwidth]{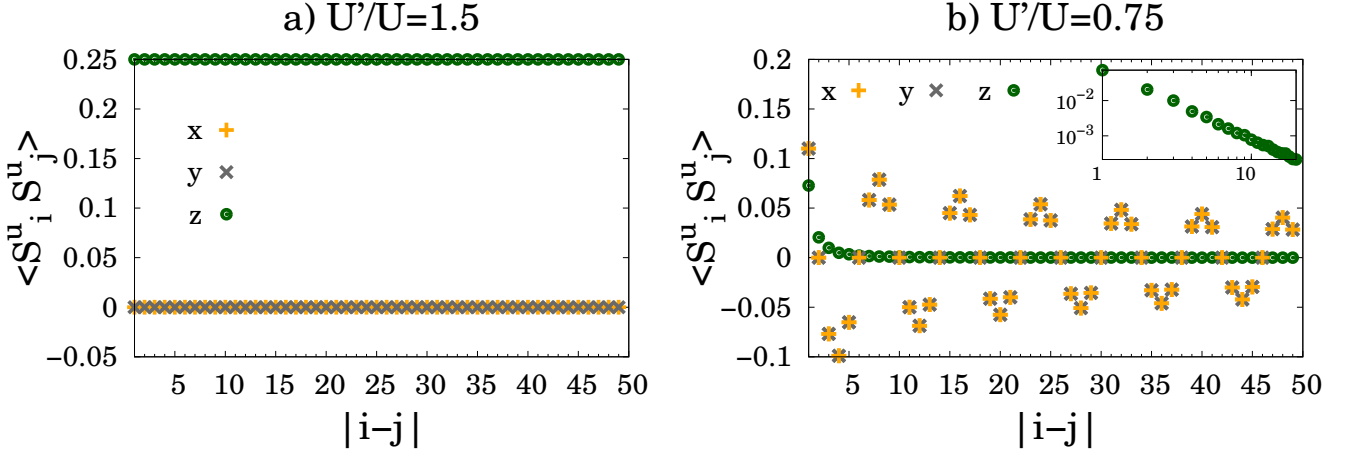}
\end{center}
\vspace{-0.2cm}
\caption{
\label{fig:correls_hof}
(Color online) Correlation functions in the ground state of Hamiltonian~(\ref{eq:Ham_Hof_B}). Shown are $\langle \Sx_i \Sx_j\rangle$ (orange plus signs), $\langle \Sy_i \Sy_j\rangle$ (gray crosses) and $\langle \Sz_i \Sz_j\rangle$ (green dots) for $\beta=\pi/8$ and (a) $U'/U=1.5$ and (b) $U'/U=0.75$.
The inset of (b) shows $\langle \Sy_i \Sy_j\rangle$ in log/log scale.
}
\end{figure*}
%-----------------------------------------%
\section{Model and Hamiltonian}
We consider an interacting two-component gas of bosons in a
one-dimensional lattice, subject to a spin-dependent artificial
magnetic field and  a synthetic spin-orbit coupling of Rashba type,
described by the Hamiltonian \be \label{eq:Ham} \Ham=
\Ham_\textbf{\tiny t}+ \Ham_\textbf{\tiny SOC} +\Ham_\textbf{\tiny
U}. \ee The kinetic part $\Ham_\textbf{\tiny t}+ \Ham_\textbf{\tiny
SOC}$ reads
\begin{align}
\Ham_\textbf{\tiny t}&=-t \cos\alpha \sum_{j} \left[ b^\dagger_{j+1,\uparrow} b_{j,\uparrow} e^{\i \beta} + b^\dagger_{j+1,\downarrow} b_{j,\downarrow}  e^{-\i \beta} \right] + \textrm{h.c.} \nonumber\\
\Ham_\textbf{\tiny SOC}&=-t \sin\alpha \sum_{j} \left[ b^\dagger_{j+1,\uparrow} b_{j,\downarrow} - b^\dagger_{j+1,\downarrow} b_{j,\uparrow} \right] + \textrm{h.c.} \, ,
\end{align}
where $t$ is the hopping amplitude,
$b_{j,\uparrow \downarrow}$ is the bosonic annihilation operator, $j$ is the site index, and $\uparrow, \downarrow$ denotes the two bosonic species (`spin' degree of freedom), $\beta$ represents the strength of a spin-dependent magnetic field, where the different bosonic species feel opposite magnetic fields, and $\alpha$ denotes the strength of the 1D Rashba spin-orbit coupling, which allows spin-flipping tunneling.
The interaction part of the Hamiltonian reads
\begin{align}
\Ham_\textbf{\tiny U}= \frac{U}{2} \sum_{j\sigma} \left[
n_{j,\sigma} (n_{j,\sigma} -1) \right] + U'  \sum_j n_{j,\uparrow}
n_{j,\downarrow}\, ,
\end{align}
where $\sigma=\uparrow,\downarrow$, and $U$ (respectively $U'$)
represents the strength of the intra- (resp. inter-) species
interaction.

When the interactions are strong enough to drive the system at unit filling into a Mott insulating phase, charge fluctuations are suppressed, and the physics is captured by an effective magnetic model.
Using the %following 
spin-1/2 representation~\cite{duan2003}
 $\Sz_j=n_{j,\uparrow}-n_{j,\downarrow}$,
  $\Sx_j=b^\dagger_{j,\uparrow} b_{j,\downarrow} + b^\dagger_{j,\downarrow} b_{j,\uparrow}$
  and $\Sy_j=-i(b^\dagger_{j,\uparrow} b_{j,\downarrow} - b^\dagger_{j,\downarrow} b_{j,\uparrow})$,
the leading terms of the effective super-exchange Hamiltonian can be derived as:
\be \label{eq:SupExHam}
\Ham_\textbf{\tiny S}= \sum_j  \left[ \sum_{u=x,y,z} J_u \Su_j \Su_{j+1} + \vec{D} \cdot (\vec{S}_j \times \vec{S}_{j+1}) \right] \, .
\ee
The Heisenberg terms are anisotropic in all  three directions (XYZ model):
\begin{align*}
J_x& = J_0 [ \sin^2\alpha - \cos^2\alpha \, \cos(2\beta) ];\\
J_y& =- J_0 [\sin^2\alpha + \cos^2\alpha \, \cos(2\beta)];\\
J_z& =J_0 (-2U'/U+1) [\cos^2\alpha-\sin^2\alpha],
\end{align*}
with $J_0=4t^2/U'$.
The parameter $U'/U$ characterizes the asymmetry of the interactions, and $U'/U=1$ represents SU(2) symmetric interactions in spin space.
The Dzyaloshinskii-Moriya (DM) interaction~\cite{dzyaloshinsky1958,moriya1960} is characterized by a three-dimensional vector  $\vec{D}$ with:
\begin{align*}
D_x&= J_0 \frac{U'}{U} \sin(2\alpha)\sin\beta; \\
D_y&= J_0 \frac{U'}{U} \sin(2\alpha) \, \cos\beta; \\
 D_z&= J_0 \cos^2\alpha \, \sin(2\beta)\, .
\end{align*}
Although there are only three independent parameters $\alpha$, $\beta$ and $U'/U$,
the effective magnetic model given in Eq.~(\ref{eq:SupExHam}) is one of the most general forms for a 1D spin-1/2 quantum magnetic model with two-site nearest-neighbor interactions.

\section{Bose-Hubbard Hamiltonian with spin-dependent magnetic field \label{sec:Hofstadter}}
Let us first focus on a relatively simple case in which only an
abelian synthetic gauge field  is present (i.e. $\alpha=0$.
This case is directly relevant to current experiments with ultra-cold
atoms~\cite{aidelsburger2013,miyake2013,atala2014}. If the different
(spin) species experience the same magnetic field, the magnetic field
has no effect on the superexchange magnetic Hamiltonian; we
thus focus on the case in which spin-$\uparrow$ and $\downarrow$
bosons feel an equal and opposite magnetic field. For $\alpha=0$,
Eq.~(\ref{eq:SupExHam}) reduces to an anisotropic Heisenberg model
(XXZ) with a DM interaction along the $z$-direction:
\begin{align} \label{eq:Ham_Hof_B}
& \Ham_\textbf{\tiny S}= \Jo \sum_j \Big[ -  \cos(2\beta) \, \left( \Sx_j \Sx_{j+1}+ \Sy_j \Sy_{j+1}\right)\\
&+ \left(-2 \frac{U'}{U}+1\right) \Sz_j \Sz_{j+1}
+ \sin(2\beta) (\Sx_j\Sy_{j+1}-\Sy_j\Sx_{j+1})\Big] . \nonumber
\end{align}

Since the anisotropy of the Heisenberg model and the DM vector are
along the same direction, the DM interaction can be gauged
away by performing a rotation of the local spin basis for
$\vec{S}_j$ around the $z$-axis by an angle $2 j \beta$
 : $\Sx_j=\cos(2 j \beta) \Spx_j+\sin(2 j \beta) \Spy_j$, $\Sy_j=\cos(2 j \beta) \Spy_j-\sin(2 j \beta) \Spx_j$ and $\Sz_j=\Spz_j$ , which leads to an XXZ model without DM interactions~\cite{brockmann2013}
\begin{align}
\Ham_\textbf{\tiny S}'' = J_0 \sum_j \Big[ - &\left( \Spx_j \Spx_{j+1} + \Spy_j \Spy_{j+1} \right) \\
&+ \left(1- 2\frac{U'}{U}\right) \, \Spz_j \Spz_{j+1} \Big] . \nonumber
\end{align}
The phase diagram of this model has been thoroughly investigated~\cite{schollwock2004,sachdev2011}.
As a result, the ground state of Hamiltonian~(\ref{eq:Ham_Hof_B}) will exhibit a gapped ferromagnetic (FM) state polarized in the $z$-direction for $U'/U >1$.
This is illustrated in Fig.~\ref{fig:correls_hof}(a), where we show the spin-spin correlation functions $\langle \Su_i \Su_j\rangle$ (with $u \in \{ x,y,z \}$) in the ground state of Eq.~(\ref{eq:Ham_Hof_B}) for  $\beta=\pi/8$ and $U'/U=1.5$.
For $0<U'/U<1$ the ground state of Hamiltonian~(\ref{eq:Ham_Hof_B}) is a gapless phase that follows from the XY phase of the XXZ model, with algebraically decaying correlations.
However, due to the rotation from the mapping, the correlations in the $x$ and $y$-directions exhibit spiral order, with a period of $\pi/\beta$ sites.
This is shown in Fig.~\ref{fig:correls_hof}(b) where the spin-spin correlations are plotted for $\beta=\pi/8$ and $U'/U=0.75$:
$\langle \Sx_i \Sx_j\rangle$ and $\langle \Sy_i \Sy_j\rangle$ oscillate with a period of 8 sites.

\section{Bose-Hubbard Hamiltonian with Rashba SOC}
%-----------------------------------------%
\begin{figure}[ht]
\begin{center}
\includegraphics[width=0.5\textwidth]{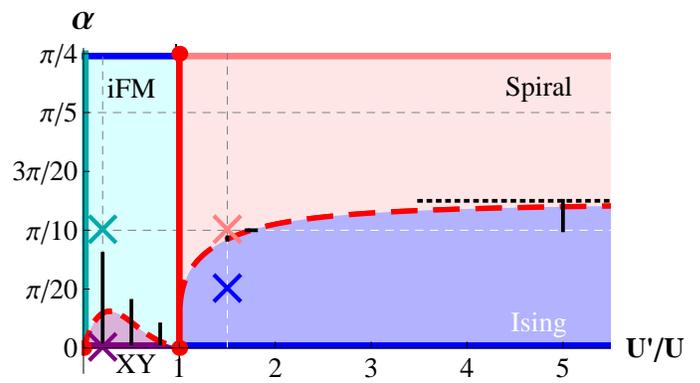}
\end{center}
\vspace{0.1cm}
\caption{
\label{fig:DiagPhase}
(Color online) Phase diagram of Hamiltonian~(\ref{eq:Ham_Rash_B}) as a function of parameters $U'/U$ and $\alpha$ obtained by DMRG calculations of the ground state.
We distinguish two gapped phases: an \emph{Ising} phase (dark blue) and one with incomplete ferromagnetic order (\emph{iFM}) (light blue); and two gapless phases: an \emph{XY phase} (purple) and a \emph{spiral} phase (pink).
The plain lines indicate when the model maps onto a XXZ model, in which case the phase is known a priori.
The colored crosses locate the points which are analyzed in Figs.~\ref{fig:phases} and \ref{fig:phases_2}, and the dotted gray lines locate the cuts analyzed in Figs.~\ref{fig:cuts1}, \ref{fig:cuts2} and \ref{fig:BKT}.
The dashed black line indicate the transition point in the $U'/U \to \infty$ limit, known analytically.
The red lines indicate critical lines: the dashed red lines are estimations of the phase boundary obtained by DMRG calculations (error bars are displayed in black) and the boundary at $U'/U=1$ is known a priori (see text).
}
\end{figure}
%-----------------------------------------%
We now study the case of a non-abelian synthetic gauge field. We consider the Rashba type of spin-orbit coupling in 1D, which
has been implemented experimentally for atoms in the continuum~\cite{stanescu2008,lin2011}.
Proposals for schemes to implement it on lattices also exist~\cite{osterloh2005,dalibard2011}.
In the case $\beta=0$, equation~(\ref{eq:SupExHam}) simplifies to:

\begin{align} \label{eq:Ham_Rash_B}
\Ham_\textbf{\tiny S}&=\Jo \sum_j \Big[  - \cos(2\alpha) \, \Sx_j \Sx_{j+1} - \Sy_j \Sy_{j+1} \nonumber \\
&+ \left(-2 \frac{U'}{U}+1\right) \cos(2\alpha) \, \Sz_j \Sz_{j+1} \nonumber \\
&+ \frac{U'}{U} \sin(2\alpha) (\Sz_j\Sx_{j+1}-\Sx_j\Sz_{j+1}) \Big] .
\end{align}
The phase diagram of this model is presented in Fig.~\ref{fig:DiagPhase} and will be explained in detail in the remainder of the paper.
We only consider the region $0 < \alpha < \pi/4$, since the rest can be deduced by simple transformations.
Indeed, Eq.~(\ref{eq:Ham_Rash_B}) is $\pi$-periodic in $\alpha$, and if $\pi/4 < \alpha < \pi/2$, by setting $\alpha'=\pi/2-\alpha$, $\Sx_j=(-1)^j\Spx_j$, $\Sy_j=\Spy_j$ and $\Sz_j=(-1)^j\Spz_j$, we recover Hamiltonian~(\ref{eq:Ham_Rash_B}) with $0 < \alpha' < \pi/4$.
In some particular cases (signaled by plain lines in Fig.~\ref{fig:DiagPhase}), Eq.~(\ref{eq:Ham_Rash_B}) can be mapped onto a XXZ model, in which case the phases are known ab-initio, as in Sec.~\ref{sec:Hofstadter}.
However, for general values of $U'/U$ and $\alpha$, the DM term cannot be gauged away, and this model can hardly be handled analytically.
We therefore explore the phase diagram numerically by computing the ground state of Hamiltonian~(\ref{eq:Ham_Rash_B}) by the Density Matrix Renormalization Group (DMRG) method~\cite{white1992,schollwock2005}.
In the calculations we use a finite-size DMRG algorithm, for systems of total sizes up to $L=500$ lattice sites with open boundary conditions.
We keep up to $m=1000$ states in the matrix product state representation.
Once the calculations are converged, the truncation error of the reduced density matrix is typically $10^{-8}$ and the energies are converged up to the $10^{\textrm{th}}$ digit.

%-----------------------------------------%
\begin{figure*}[ht]
\begin{center}
\includegraphics[width=1.0\textwidth]{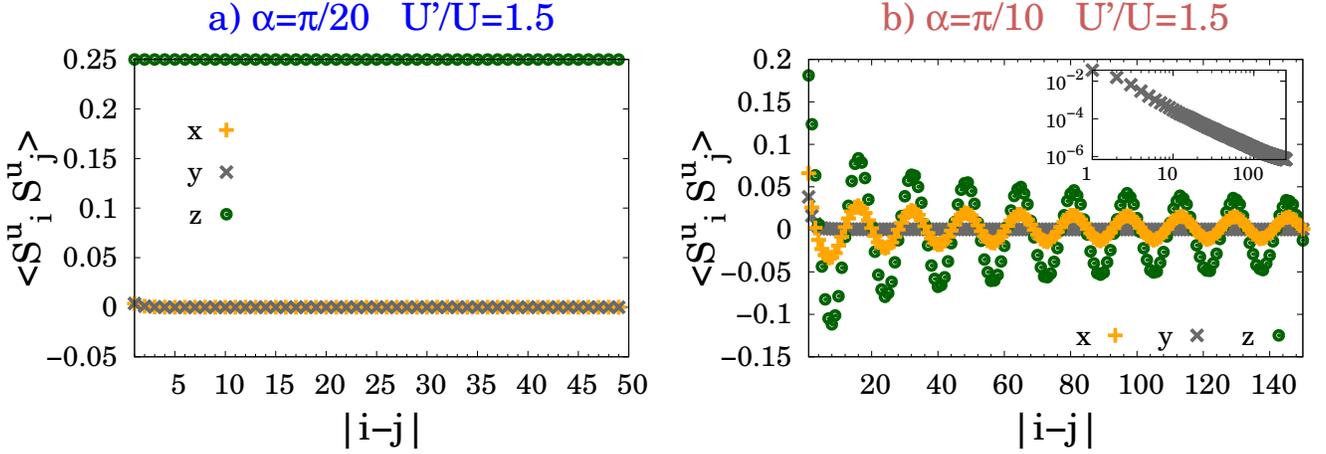}
\end{center}
\vspace{-0.2cm}
\caption{
\label{fig:phases}
(Color online) Correlation functions in the ground state  of Hamiltonian~(\ref{eq:Ham_Rash_B}) obtained by DMRG calculations.
%~\cite{note_dmrg}. 
Shown are $\langle \Sx_i \Sx_j\rangle$ (orange plus signs), $\langle \Sy_i \Sy_j\rangle$ (gray crosses) and $\langle \Sz_i \Sz_j\rangle$ (green dots) for (a) $\alpha=\pi/20$ and $U'/U=1.5$ and (b) $\alpha=\pi/10$ and $U'/U=1.5$.
The inset of (b) show $\langle \Sy_i \Sy_j\rangle$ in log/log scale.
For the sake of clarity those points are identified in Fig.~\ref{fig:DiagPhase} by crosses.
}
\end{figure*}
%-----------------------------------------%
\subsection{$U'>U$ -- Ising to spiral phase transition \label{Ising-spiral}}
Let us first analyze the region $U'>U$ of the phase diagram,
where two different phases exist. The spin-spin correlation functions $\langle \Su_i \Su_j\rangle$  (with $u \in \{x,y,z\}$) of these two phases in the ground state are presented in figure~\ref{fig:phases}.
To get some insight, we first focus on several special points (lines) in the phase diagram.

Firstly, along the $\alpha=0$ axis, Eq.~(\ref{eq:Ham_Rash_B}) becomes a XXZ model:
\begin{align} \label{eq:Ham_alpha0}
\Ham_\textbf{\tiny S}^{\alpha=0}=& \Jo \sum_j \Big[  - \left( \Sx_j \Sx_{j+1} + \Sy_j \Sy_{j+1} \right) \\
&+ \left(-2 \frac{U'}{U}+1\right) \Sz_j \Sz_{j+1} \Big]. \nonumber
\end{align}
For $U'/U>1$ the ground state is therefore an Ising state with
perfect FM ordering along the $z$-direction~\cite{schollwock2004}. For small values of $\alpha$, the
term $-\Sz_j \Sz_{j+1}$ still dominates in $\Ham_\textbf{\tiny S}$
and it is easy to prove that the ground state is still a nearly-perfect
FM phase (a gapped Ising-type phase) as shown in the dark
blue region in phase diagram of Fig.~\ref{fig:DiagPhase}. This can
be numerically verified by the spin-spin correlation functions as
shown in Fig.~\ref{fig:phases}(a), where we can observe that in this
regime the ground state exhibits nearly-perfect FM order in the
$z$-direction with $\langle\Sz_i \Sz_j\rangle \simeq 1/4$ for any
$|i-j|$.

Secondly, if we focus on the line $\alpha = \pi/4$, the Hamiltonian of Eq.~(\ref{eq:Ham_Rash_B})
is given by
\begin{align} \label{eq:Ham_alpha0-25}
\Ham_\textbf{\tiny S}^{\alpha=\pi/4} =\Jo \sum_j \left[ - \Sy_j \Sy_{j+1} + \frac{U'}{U} (\Sz_j\Sx_{j+1}-\Sx_j\Sz_{j+1}) \right].
\end{align}
After  a rotation of the local basis of each spin $\vec{S}_i$ by an
angle $j\pi/2$ around axis $y$ this maps onto $\Ham_\textbf{\tiny S}''^{ \alpha=\pi/4}=\Jo \sum_j  -
\Spy_j \Spy_{j+1} + \frac{U'}{U} (\Spx_j \Spx_{j+1}+\Spz_j \Spz_j)$.
Therefore, for $U'/U>1$, the system
is in a gapless `XY phase'~\cite{schollwock2004}
(with uniaxial symmetry around the $y$-axis). It features
algebraic decay of the correlations, and a spiral order with a
4-site period along $y$, due to the rotation of the mapping. This
picture does not qualitatively change when $\alpha$ is close to
$\pi/4$, in which case the DM term dominates in
Eq.~(\ref{eq:Ham_Rash_B}), and we find a spiral phase around the $y$-direction in the region of Fig.~\ref{fig:DiagPhase} shaded in pink.
The correlations also decay algebraically [see
Fig.~\ref{fig:phases}(b)], signaling a gapless Luttinger-liquid
phase~\cite{cazalilla2011}. Moreover, we find that the correlations
$\langle \Su_i \Su_j\rangle$ in the $x$ and $z$-directions oscillate
with the same period, thus showing spiral order~\cite{note_exps}\nocite{imriska2014}. Note that the spiral
does not have the same amplitude in the $x$ and $z$-directions, due
to the anisotropy in the exchange term in Eq.~(\ref{eq:Ham_Rash_B}).

In the limit $U'/U\rightarrow \infty$, the Hamiltonian in Eq.~(\ref{eq:Ham_Rash_B}) reduces to
\begin{align} \label{eq:Ham_Infty}
\Ham_\textbf{\tiny S}^{\frac{U'}{U} \to \infty} & \simeq \Jo \frac{U'}{U} \sum_j \Big[ -2 \cos(2\alpha) \, \Sz_j \Sz_{j+1} \\
&+ \sin(2\alpha) (\Sz_j\Sx_{j+1}-\Sx_j\Sz_{j+1}) \Big], \nonumber
\end{align}
which can be solved exactly by performing a rotation in spin space
followed by the Jordan-Wigner
transformation~\cite{jordan1928,cazalilla2011}.
Equation~(\ref{eq:Ham_Infty}) can be mapped to a % (quasi-)
noninteracting spinless fermion Hamiltonian, with the dispersion
relation:
\be \epsilon^\pm_k = \Jo \frac{U'}{U} \left[
\frac{\tan(2\alpha)}{2} \sin(k) \pm \frac{1}{2} \right],
\ee
where $k$ is the wavevector in units of the reciprocal lattice spacing. One
immediately finds a phase transition from a gapped phase 
%at small values of $\alpha$ and 
to a gapless phase 
%for higher values, 
with increasing $\alpha$,
with the phase transition taking place at $\alpha_\textbf{\tiny c}=\pi/8$. This value
is represented as a dotted black line in Fig.~\ref{fig:DiagPhase}.

%-----------------------------------------%
\begin{figure}[ht]
\begin{center}
\includegraphics[width=0.5\textwidth]{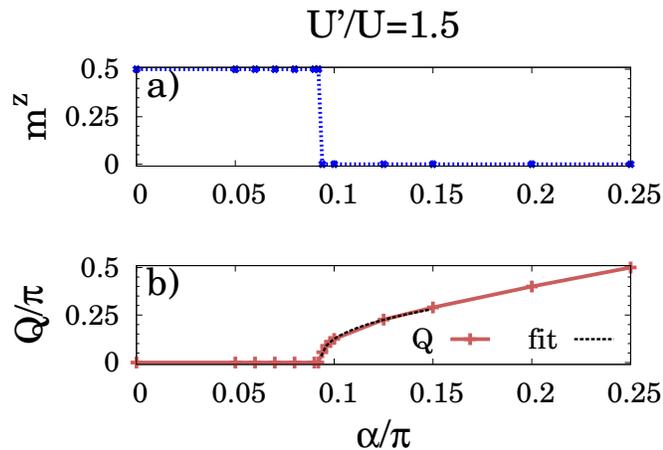}
\end{center}
\vspace{-0.2cm}
\caption{
\label{fig:cuts1}
(Color online) (a) Magnetization $m^{z}$ in the $z$-direction and (b) ordering wavevector $Q$
at constant $U'/U=1.5$.
This cut is highlighted in Fig.~\ref{fig:DiagPhase} by a dashed gray line.
The magnetization is extracted from correlation functions in the ground state with $m^{z}=\sqrt{\lim_{|i-j| \to \infty} \langle S^{z}_i S^{z}_j\rangle}$ and $Q$ from large-distance fits of $\langle\Sx_i \Sx_j\rangle$ and $\langle\Sz_i \Sz_j\rangle$ by $\cos(Q |i-j|)/|i-j|^\gamma$ with $Q$ and $\gamma$ as fitting parameters.
In (b) the data near the transition point is fitted by $Q \propto (\alpha-\alpha_\textbf{\tiny c})^\delta$ with $\delta$ a fitting parameter (black dashed line). The fit gives $\alpha_\textbf{\tiny c}=0.093 \pi$ and $\delta=0.38$.
}
\end{figure}
%-----------------------------------------%
For finite values of $U'/U$, the problem can no longer been solved analytically
We have thus computed the ground state properties with DMRG.
The magnetization $m^z=\sqrt{\lim_{|i-j| \to \infty} \langle\Sz_i \Sz_j\rangle}$ is plotted  in Fig.~\ref{fig:cuts1}(a) as a function of $\alpha$ for fixed $U'/U=1.5$.
This shows that for small $\alpha$
the ground state is a nearly-perfect ferromagnet. % (the system is in a gapped Ising phase).
At the transition point  [$\alpha_\textbf{\tiny c} \simeq (0.093 \pm 0.001)\pi$], the magnetization suddenly drops to $0$, therefore signaling a first-order phase transition.

Another way to characterize the phase transition is to compute the
characteristic wavevector $Q$ of the  spiral phase, in which the
correlations become incommensurate. The incommensurability leads to
a shift of the peak in the structure factor $\mathcal{S}_a(q)
=\langle \sum_{i,j} e^{\i q (i-j)} \Su_i \Su_j \rangle/L$, with $L$
the system size, from the position $q=0$ to $q=\pm Q$, where the
peaks are broadened by the decay of correlations and possibly
finite-size effects. We obtain the characteristic wavevector $Q$ by
fitting $\langle\Sx_i \Sx_j\rangle$ and $\langle\Sz_i \Sz_j\rangle$
by $\cos(Q |i-j|)/|i-j|^\gamma$ with $Q$ and $\gamma$ as fitting parameters.
The characteristic wavevector $Q$ as
a function of $\alpha$ is shown in Fig.~\ref{fig:cuts1}(b). We
find that $Q$ increases continuously from zero when crossing the
critical point.
At the transition, it exhibits a singularity compatible with
$Q(\alpha)\propto (\alpha-\alpha_\textbf{\tiny c})^\delta$ as can be
seen from the fit in Fig.~\ref{fig:cuts1}(b). At $\alpha=\pi/4$, $Q$
reaches $\pi/2$, which agrees with our previous analysis [see
Eq.~(\ref{eq:Ham_alpha0-25}) and below].

\begin{figure*}[ht]
\begin{center}
\includegraphics[width=1\textwidth]{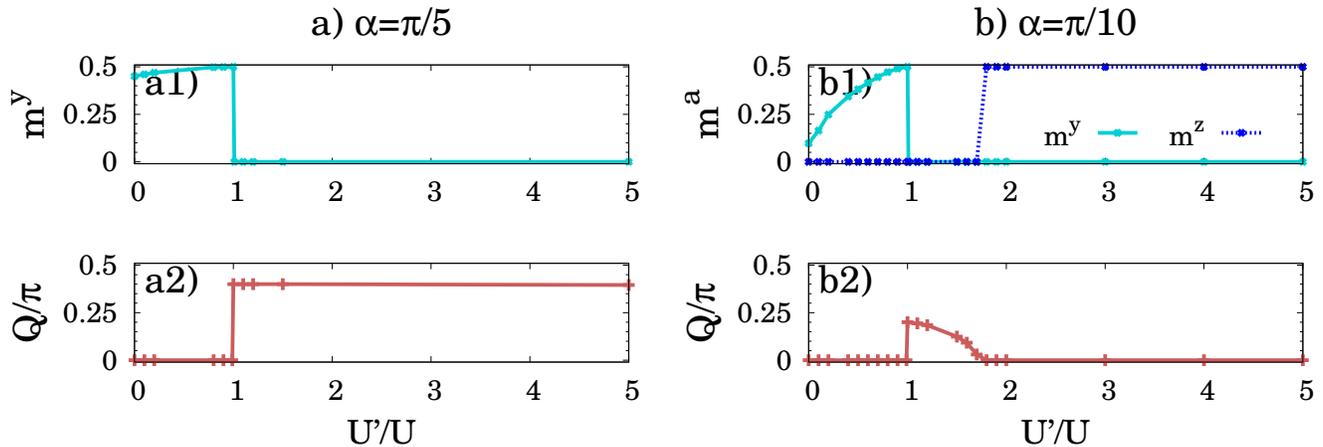}
\end{center}
\vspace{-0.2cm}
\caption{
\label{fig:cuts2}
(Color online) Magnetization $m^{y,z}$ in the $y$ and $z$-directions and ordering wavevector $Q$
along two cuts in the phase diagram: (a) constant $\alpha=\pi/5$ and (b) constant $\alpha=\pi/10$.
Those cuts are highlighted in Fig.~\ref{fig:DiagPhase} by dashed gray lines.
The magnetization is extracted from correlation functions in the ground state with $m^{y,z}=\sqrt{\lim_{|i-j| \to \infty} \langle S^{y,z}_i S^{y,z}_j\rangle}$ and $Q$ from large-distance fits of $\langle\Sx_i \Sx_j\rangle$ and $\langle\Sz_i \Sz_j\rangle$ by $\cos(Q |i-j|)/|i-j|^\gamma$.
}
\end{figure*}

The critical value $\alpha_\textbf{\tiny c}$ obtained above for
$U'/U=1.5$ is marked in Fig.~\ref{fig:DiagPhase},  together with
an estimation of the transition point along the lines $U'/U=5$ and
$\alpha=\pi/10$ [see also Fig.~\ref{fig:cuts2}(b) below].
The critical line for this transition must eventually bend down
to reach  the point $(\alpha, U'/U)=(0,1)$, because the line
$U'/U=1$ is a spiral critical phase (see below); hence the
phase diagram for $U'/U>1$ as shown in Fig.~\ref{fig:DiagPhase}.

%-----------------------------------------%
%-----------------------------------------%
\subsection{$U' \simeq U$ -- Incomplete ferromagnet to spiral phase transition}

Now, we focus on the regime $U' \simeq U$.
%  and focus on the phase transition at $U'/U=1$.
We again start by analyzing the special cases
of the phase diagram to get some intuition.

In the case $U'=U$, Hamiltonian~(\ref{eq:Ham_Rash_B}) reads
\begin{align} \label{eq:Ham_1}
\Ham_\textbf{\tiny S}^{\frac{U'}{U}=1} &=\Jo \sum_j  \Big[ - \cos(2\alpha) \, \left( \Sx_j \Sx_{j+1} + \Sz_j \Sz_{j+1} \right) - \Sy_j \Sy_{j+1} \nonumber \\
&+ \sin(2\alpha) (\Sz_j\Sx_{j+1}-\Sx_j\Sz_{j+1}) \Big],
\end{align}
and a rotation of the local basis of each  spin $\vec{S}_i$ by an
angle $2j\alpha$ around axis $y$ allows us to gauge out the DM
term~\cite{cai2012}. The model then reduces to an isotropic
FM Heisenberg model $\Ham_\textbf{\tiny
S}''^{\frac{U'}{U}=1}=- {\vec{S}}'_j \cdot {\vec{S}}'_{j+1}$. In
that case, the ground state is a gapless FM state
with high degeneracy, which is also the critical state in the XXZ
model~\cite{schollwock2004}.

%-----------------------------------------%
\begin{figure*}[ht]
\begin{center}
\includegraphics[width=1.0\textwidth]{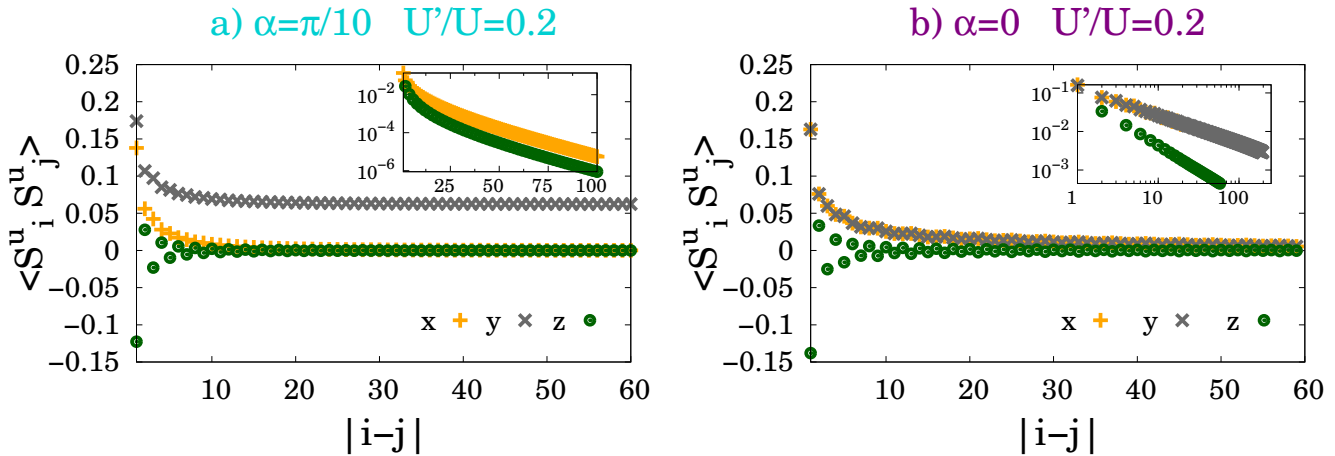}
\end{center}
\vspace{-0.2cm} \caption{ \label{fig:phases_2}
(Color online) Correlation functions
in the ground state of  Hamiltonian~(\ref{eq:Ham_Rash_B}) obtained by DMRG calculations.
%~\cite{note_dmrg}. 
Shown are $\langle \Sx_i \Sx_j\rangle$ (orange plus signs), $\langle \Sy_i
\Sy_j\rangle$ (gray crosses) and $\langle \Sz_i \Sz_j\rangle$ (green
dots) for (a) $\alpha=\pi/10$ and $U'/U=0.2$ and (b) $\alpha=0$ and
$U'/U=0.2$. The insets show the same data in log/linear scale for
(a), and in log/log scale for (b). For the sake of clarity those
points are identified in Fig.~\ref{fig:DiagPhase} by crosses. }
\end{figure*}
%-----------------------------------------%
Figure~\ref{fig:cuts2} displays $m^y=\sqrt{\lim_{|i-j| \to \infty}
\langle\Sy_i \Sy_j\rangle}$, $m^z$ and  the characteristic
wavevector $Q$ in the spiral phase as a function of $U'/U$ with
fixed $\alpha=\pi/5$ [Fig.~\ref{fig:cuts2} (a)]   and $\pi/10$
[Fig.~\ref{fig:cuts2} (b)] respectively. In both cases we find that
$U'/U=1$ is indeed the critical point of a first-order transition
between a gapless spiral phase and a gapped phase with long-range
FM order along the $y$-direction ($m^y \neq 0$).
We name this phase `incomplete ferromagnet' (iFM) in the following, since $m^y <1/4$.
It is found in the top-left corner of the phase diagram
(blue-shaded region in Fig.~\ref{fig:DiagPhase}), a region in which
the Hamiltonian is dominated by $-\Sy_j \Sy_{j+1}$.
Figure~\ref{fig:phases_2}(a) shows the spin-spin correlation
functions at $U'/U=0.2$ and $\alpha=\pi/10$. We indeed find a
reduced magnetization in the $y$-direction: $\lim_{|i-j| \to \infty}
\langle\Sy_i \Sy_j\rangle \simeq 0.062 < 1/4$. We also observe that
the correlations along the $x$ and $z$-directions decay
exponentially, indicating a gapped phase. The nature of the iFM
state can be understood in the limit $U'/U=0$, where
Eq.~(\ref{eq:Ham_Rash_B}) becomes
\begin{align} \label{eq:Ham_0}
\Ham_\textbf{\tiny S}^{\frac{U'}{U}=0}  &=\Jo \sum_j  \left[ \cos(2\alpha) \left( - \Sx_j \Sx_{j+1}+ \Sz_j \Sz_{j+1} \right) - \Sy_j \Sy_{j+1} \right].
\end{align}
After the rotation of each spin $\vec{S}_j$ around the $z$-axis by an angle $j \pi$
[$\Sx_j=(-1)^j\Spx_{j+1}$, $\Sy_j=(-1)^j\Spy_j$ and $\Sz_j=\Spz_j$],
the model maps to an antiferromagnetic (AF) XXZ model in the 
N\'eel phase
%easy-axis phase 
(with preferred axis $y$).
Unlike for the
FM XXZ model, it is known that the perfect AF state along the $y$-axis
(which would correspond to a perfect FM state along the $y$-direction after the above rotation)
is not the ground state of the AF XXZ model in the N\'eel phase.
% easy-axis phase.
Therefore, after the above spin rotation,  the ground state of
Eq.~(\ref{eq:Ham_0}) exhibits FM order with reduced
magnetization along the $y$-axis.

We therefore find that the transition from the iFM phase to the incommensurate spiral phase is again a first-order phase transition.
As can be seen in Figs.~\ref{fig:cuts2}(a2) and (b2), the onset of incommensurability also coincides with the transition point.
However, unlike the Ising to spiral phase transition analyzed previously (see Sec.~\ref{Ising-spiral}), we find that the characteristic wavevector jumps discontinuously from 0 (in the commensurate iFM phase) to a finite value $Q=2\alpha$.
In Fig.~\ref{fig:cuts2}(a2) we observe that the value of $Q$ is rather constant in the spiral phase, along a cut at $\alpha=\pi/5$.
In the cut at $\alpha=\pi/10$ [Fig.~\ref{fig:cuts2}(b2)] however, $Q$ decreases and reaches 0 at $U'/U \simeq 1.75 \pm 0.01$, at the point where the magnetization $m^z$ jumps to $0.5$, indicating a transition to the Ising phase (nearly-perfect FM phase).

%-----------------------------------------%
\begin{figure}[ht]
\begin{center}
\includegraphics[width=0.5\textwidth]{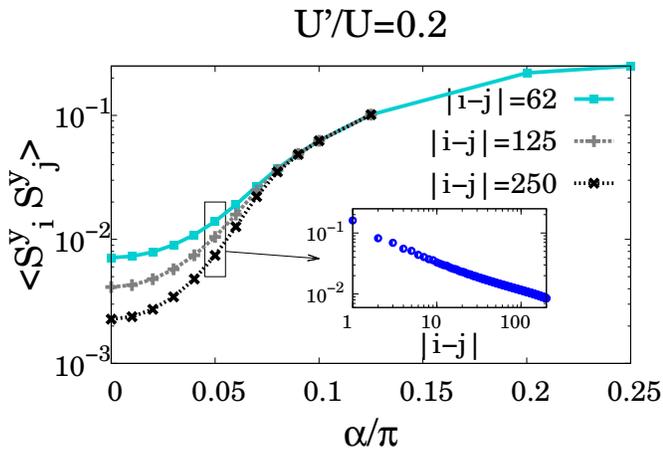}
\end{center}
\vspace{0.1cm}
\caption{
\label{fig:BKT}
(Color online) Spin-spin correlations $\langle\Sy_{i} \Sy_{j}\rangle$, at constant $U'/U=0.2$, in logarithmic scale.
They are computed in the ground state obtained by DMRG,
%~\cite{note_dmrg}, 
around the center of a system of total size $L=500$, for different site distances: $|i-j|=50$ (plain blue line), $125$ (dashed grey line) and $250$ (dotted black line).
This cut is highlighted in Fig.~\ref{fig:DiagPhase} by a dashed gray line.
The inset shows the full function $\langle\Sy_{i} \Sy_{j}\rangle$ for $\alpha=\pi/20$, in log/log scale.
}
\end{figure}
%-----------------------------------------%
\subsection{$U' < U$ -- Incomplete ferromagnet to XY phase transition}
Finally, we study the quantum phases  and transitions in  the region $U'<U$, and again first focus on some special points or lines in the phase diagram to get some insight.

We note firstly that in  the limit $\alpha=0$ [see Eq.~(\ref{eq:Ham_alpha0})], the
ground state for $0<U'/U<1$ is a gapless XY phase~\cite{schollwock2004}.
% (easy-plane spin state within $xy-$plane.
In this phase, all correlation functions
decay algebraically, as shown in Fig.~\ref{fig:phases_2}(b) for
$U'/U=0.2$ and $\alpha=0$. We expect that a small $\alpha$ will not
qualitatively change the nature of this gapless phase. Therefore,
there should be a phase transition to the gapped iFM phase with
increasing $\alpha$. Such a gapless phase with respect to spin
excitations is also predicted in Ref.~\cite{zhao2013}.

Let us focus on this phase transition. Along the $U'/U=0$ line, as
we have analyzed above, the Hamiltonian maps onto an AF XXZ model
[see Eq.~(\ref{eq:Ham_0}) and below], and the ground state is always
of gapped iFM type for all $\alpha>0$, while the point at
$(\alpha,U'/U)=(0,0)$ is a critical point of
Berezinskii-Kosterlitz-Thouless (BKT)
type~\cite{sachdev2011,schollwock2004}.
Therefore, for small
$\alpha$, the excitation gap is exponentially small, and varies as $\Delta \sim
e^{-\pi/(\sqrt{2} \alpha)}$~\cite{sachdev2011,schollwock2004}, which
makes it very difficult to numerically distinguish the gapless
from the gapped phase.

In figure~\ref{fig:BKT}, we show the correlation $\langle \Sy_i
\Sy_j\rangle$ as a function of $\alpha$ for $U'/U=0.2$ and 
various site separations $|i-j|$. For large $\alpha$ (e.g.
$\alpha>\pi/10$), spin-spin correlations along the $y$-direction
saturate to a non-zero value at large distances, which indicates the
existence of FM long range ordering, while for smaller $\alpha$
(e.g. $\alpha=\pi/20$),  the spin-spin correlations decay
algebraically at large distances (as shown in the inset of
Fig.~\ref{fig:BKT}), which seems to indicate a gapless phase.
However, due to the numerical precision and limitations in the system size
of our numerical calculations,  we cannot exclude the possibility
that it is a gapped phase with an extremely small gap, just as the
situation near the BKT phase transition.
In which case the spin correlations
would saturate to an extremely small value.
It is therefore difficult to locate the position of the assumed transition point precisely.
However, our data clearly provide an upper bound for the phase
transition point.
Indeed, the ground state at $\alpha=\pi/10$ has
finite magnetization, and the other correlation functions $\langle
\Sx_i \Sx_j\rangle$ and $\langle \Sz_i \Sz_j\rangle$ decrease
exponentially with $|i-j|$ [see Fig.~\ref{fig:phases_2}(a)].
We confirmed using infinite-size DMRG that the correlation length 
saturates in the thermodynamic limit to a finite value
[we find $\xi = 21.16$ lattice sites for the parameters of Fig.~\ref{fig:phases_2}(a)], thus
demonstrating unambiguously that this point is in the gapped iFM
phase. Our numerical result indicates that there is a continuous
phase transition between the iFM and the XY phase (the purple regime
in Fig.~\ref{fig:DiagPhase}). However, as we analyzed before, we
cannot preclude here the possibility that there is no phase
transition at finite $\alpha$, and that the gapless regime shrinks
to a critical line with $\alpha=0$.

\section{Conclusion and discussion}
In this paper we have studied the quantum magnetism of the Mott
insulating phases found in the strongly interacting limit of the 1D
Bose-Hubbard model with both abelian and non-abelian synthetic gauge
fields. In the abelian case (spin-dependent magnetic field) which is
relevant to current experiments with cold atoms, we found that the
ground state exhibits a spiral quasi long-range order in the regime
$U'/U<1$. In the non-abelian case (Rashba spin-orbit coupling), we have
studied the phase diagram of the effective Hamiltonian, where we
identified four phases with different magnetic textures: two gapped
phases with nearly-complete (Ising) and incomplete (iFM) ferromagnetic
order and two gapless phases with and without spiral quasi long-range
order. We have found that the transitions between the ferromagnetic
phases and the spiral phase are both first order, and the emergence
of the incommensurability in the spiral phase coincides with the phase transition.
The ordering wavevector
is continuous at the Ising to spiral phase transition, whereas it is
discontinuous at the iFM to spiral one. Finally, in the regime
$U'/U<1$, there is a continuous phase transition from an XY phase to
a gapped iFM phase, for small $\alpha$.

We have focused here on situations in which only one component of the DM interaction is non-zero.
However, the general model given in Eq.~(\ref{eq:SupExHam}) allows
more than one component for the DM vector $\vec{D}$, in the case where both
the abelian and the non-abelian synthetic gauge fields are present. % at the same time.
This may give rise to richer quantum
magnetic phases and deserves to be explored in the future.
We have also focused on the 1D case. However, DM interactions are also very interesting in 2D,
as they can give rise to exotic topological
magnetic textures such as vortex and skyrmion crystals: topics that could
also be addressed with ultracold
atoms~\cite{cai2012,radic2012,cole2012}.

\paragraph*{Note --}
During the completion of this manuscript, we became aware of the
work reported in Refs.~\cite{zhao2014,xu2014,peotta2014}, which address a similar topic.

%%%%%%%%%%%%%%%%%%%%%%%%%%%%%%%%%%%%%%%%%%%%%%%%%%%%%%%%%%%%%%%%%%%%%%

 \section{ACKNOWLEDGMENTS}
We thank Fabian Heidrich-Meisner, Vincenzo Alba and Lode
Pollet for enlightening discussions. This research was supported by
DFG FOR 801.
This work was partly supported by the Australian Research Council grant CE110001013.

%%%%%%%%%%%%%%%%%%%%%%%%%%%%%%%%%%%%%%%%%%%%%%%%%%%%%%%%%%%%%%%%%%%%%%

\end{document}